\documentclass[twocolumn]{aastex631}

\usepackage{graphicx}
\usepackage{dcolumn}
\usepackage{bm}

\usepackage{natbib}
\usepackage{latexsym} 
\usepackage{times} 
\usepackage{hyperref}
\usepackage{mathrsfs}
\usepackage[utf8x]{inputenc}
\usepackage{graphicx}
\usepackage{booktabs}
\usepackage{array}
\usepackage{multirow}
\usepackage{mathtools}
\usepackage{cancel}
\usepackage{comment}
\usepackage{float}
\usepackage[normalem]{ulem}

\newcommand{\citeAliasTwo}[2]{\defcitealias{#1}{#2}\citetalias{#1}}

\shortauthors{Cattorini et al.}

\graphicspath{{./}{figures/}}

\begin{document}
\title{Misaligned Spinning Binary Black Hole Mergers in Hot Magnetized Plasma}
\correspondingauthor{Federico Cattorini}
\email{fcattorini@uninsubria.it}

\author[0000-0002-3907-9583]{Federico Cattorini}
\affiliation{DiSAT, Università degli Studi dell'Insubria, Via Valleggio, 11, I-22100 Como , Italy}
\affiliation{INFN, Sezione di Milano-Bicocca, Piazza della Scienza 3, I-20126 Milano, Italy}

\author{Sofia Maggioni}
\affiliation{Dipartimento di Fisica G. Occhialini, Universit\`a di Milano-Bicocca, Piazza della Scienza 3, I-20126 Milano, Italy}

\author[0000-0002-6947-4023]{Bruno Giacomazzo}
\affiliation{Dipartimento di Fisica G. Occhialini, Universit\`a di Milano-Bicocca, Piazza della Scienza 3, I-20126 Milano, Italy}
\affiliation{INFN, Sezione di Milano-Bicocca, Piazza della Scienza 3, I-20126 Milano, Italy}
\affiliation{INAF, Osservatorio Astronomico di Brera, Via E. Bianchi 46, I-23807 Merate, Italy}

\author[0000-0003-3291-3704]{Francesco Haardt}
\affiliation{DiSAT, Università degli Studi dell'Insubria, Via Valleggio, 11, I-22100 Como , Italy}
\affiliation{INFN, Sezione di Milano-Bicocca, Piazza della Scienza 3, I-20126 Milano, Italy}
\affiliation{INAF, Osservatorio Astronomico di Brera, Via E. Bianchi 46, I-23807 Merate, Italy}

\author{Monica Colpi}
\affiliation{Dipartimento di Fisica G. Occhialini, Universit\`a di Milano-Bicocca, Piazza della Scienza 3, I-20126 Milano, Italy}
\affiliation{INFN, Sezione di Milano-Bicocca, Piazza della Scienza 3, I-20126 Milano, Italy}

\author{Stefano Covino}
\affiliation{INAF, Osservatorio Astronomico di Brera, Via E. Bianchi 46, I-23807 Merate, Italy}
\affiliation{DiSAT, Università degli Studi dell'Insubria, Via Valleggio, 11, I-22100 Como , Italy}
\begin{abstract}
We present general relativistic magnetohydrodynamical simulations of equal-mass spinning black hole binary mergers embedded in a magnetized gas cloud.  We focus on the effect of the spin orientation relative to the orbital angular momentum on the flow dynamics, mass accretion rate and Poynting luminosity. We find that, across the inspiral, the gas accreting onto the individual black holes concentrates into disklike overdensities, whose angular momenta are oriented towards the spin axes and which persist until merger. We identify quasiperiodic modulations occurring in the mass accretion rate at the level of $\sim$1\%-20\%, evolving in parallel with the gravitational wave chirp. The similarity between the accretion rate time-series and the gravitational strain is a consequence of the interplay between strong, dynamical gravitational fields and magnetic fields in the vicinity of the inspiralling black holes. This result suggests that quasiperiodicity in the pre-merger accretion rate of massive binaries is not exclusive of environments in which the black holes are embedded in a circumbinary accretion disk, and could provide an additional useful signature of electromagnetic emission concurrent to low-frequency gravitational-wave detection.
\end{abstract}
\section{Introduction}\label{sec:intro}
\noindent 
Most galaxies are believed to host a central massive black hole (MBH) and, as a result of galaxy mergers, MBH binaries (MBHBs) are expected to form \citep[see, e.g.,][for a review]{Colpi-2014}.
The MBHBs are understood to be the loudest sources of low-frequency gravitational waves (GW), whose detection will be one of the main scientific goals of future spaceborne interferometers such as LISA \cite[]{LISA-2017}.

Since MBHB mergers are anticipated to occur in gas-rich environments,  we expect these events to be sources of conspicuous electromagnetic (EM) radiation as well, yielding unique opportunities for multi-messenger detections \cite[][]{Bogdanovic-2021-review}. In order to understand the mechanisms which may give rise to such EM counterparts, it is crucial to confront future multi-messenger observations with magnetohydrodynamical models of the gas around spinning MBHs in proximity and after merger.
Due to the complexity of the underlying phenomena, this modeling needs to be mostly numerical.

Over the last decade, many numerical studies explored the evolution of gas around MBHBs, progressively adding the layers of physics which are needed to investigate the physical mechanisms which govern the observable signatures of a MBHB inspiral and merger. There is still large uncertainty about the type of environment found in the proximity of merging MBHs; so far, these extreme events have been examined in two limiting scenarios, i.e., the (i) circumbinary disk (CBD) and the (ii) gas cloud model. The former has been the subject of numerous numerical investigations, and has been explored by several theoretical groups with different techniques: Newtonian (and pseudo-Newtonian) viscous hydrodynamics \cite[][]{MacFadyen-2008, D'Orazio-2016, Tang-2018, Tiede-2020} and magnetohydrodynamics \cite[MHD, ][]{Shi-2012,  Shi-2016}; MHD evolutions over Post-Newtonian (PN) spacetime metrics \cite[][]{Noble-2012, Zilhao-2015, Bowen-2017, Bowen-2018, Noble-2021}; fully general relativistic magnetohydrodynamic (GRMHD) simulations \cite[][]{Farris-2012, Gold-2014a, Gold-2014b}. These works studied the CBD response to MBHB inspiral \cite[][]{Noble-2012}, the mass feeding to individual ``mini-disks'' around inspiraling MBHs \cite[][]{Bowen-2017, Bowen-2018}, and the EM radiation emerging from these systems employing ray-tracing techniques \cite[][]{D'Ascoli-2018, Gutierrez-2022}. More recently, the impact of the spins of individual BHs on the dynamics of mini-disks was explored by \cite{Paschalidis-2021} in full GR, and by \cite{Lopez-Armengol-2021}, employing an approximate metric for the spacetime evolution \citep{Combi-2021, Combi-2021-arxiv}.

If the accretion flow surrounding the binary is radiatively inefficient, a gas cloud scenario is a fair approximation of the physical conditions of matter in the vicinity of the BHs \cite[][]{Bogdanovic-2011}. In the first numerical studies of such a scenario \cite[][]{Farris-2010, Bode-2012}, the late-inspiral and  merger take place in a hot, homogeneous  cloud in which the gas is either at rest or moving relative to the binary. \cite{Giacomazzo-2012} pioneered the theoretical study of merging binaries in magnetized gas clouds using full GRMHD techniques. A later development by \cite{Kelly-2017} explored how BHB mergers are affected by different values of the gas magnetization parametrized  by the magnetic-to-gas pressure ratio $\beta^{-1} = p_{\mathrm{mag}}/p_{\mathrm{gas}}$. These works both considered merging equal-mass, non-spinning BHBs immersed in a diffuse hot gas initially threaded by a uniform magnetic field aligned with the orbital angular momentum and examined the evolution of the mass accretion rate and the development of EM energy as Poynting flux. The aftermath of the merger was further investigated by \cite{Kelly-2021}, who focused on the steady-state behavior of the magnetized gas around a post-merger Kerr black hole.

\vspace{.2cm}
In \cite{Cattorini-2021}, hereafter referred to as Paper I, we extended the analysis of \cite{Giacomazzo-2012} and \cite{ Kelly-2017} and performed the first simulations of equal-mass binaries of merging spinning BHs.
We covered a range of initially uniform, moderately magnetized fluids with different initial values of $\beta^{-1}$. For each value of $\beta^{-1}$, we analyzed distinct spin configurations defined by adimensional spin parameters $a = a_z = (0, 0.3, 0.6)$, and explored the dependence of the accretion rate $\dot{M}$ and of the Poynting luminosity $L_{\mathrm{Poynt}}$ on the magnitude of $a$. We found that, for a given initial value of $\beta^{-1}$, spin exerts a suppressing effect on the mass accretion rate; conversely, the post-merger peak Poynting luminosity of spinning BHB remnants can be enhanced by up to a factor of $\sim2.5$. 
All configurations examined in \citeAliasTwo{Cattorini-2021}{Paper I} considered binaries of spinning BHs with both spins aligned with the orbital angular momentum $L_{\mathrm{orb}}$. 
Recent work by \cite{Kelly-2021} and \cite{Ressler-2021} began to explore the effect of magnetic field orientation with respect to the spin axis of a single accreting BH, e.g., varying the angle $\theta_{\mathrm{B}}$ between the asymptotic magnetic field and the BH spin direction and investigating the sensitivity on $\theta_{\mathrm{B}}$ of the Poynting luminosity $L_{\mathrm{Poynt}}$ and the accretion rate $\dot{M}$.

\begin{table*}
\centering
\caption{BBH initial data parameters and derived quantities in code units of the GRMHD runs: initial puncture separation $a_0$ and linear momentum components $p_x$ \& $p_y$, dimensionless spin vectors $\hat{a}_i = (a_{i,x}, a_{i,y}, a_{i, z})$ of each BH, the merger time $t_{\mathrm{merger}}$, the remnant's mass M$_{\mathrm{rem}}$, the $z$-component of the remnant's spin parameter $a_{z, \mathrm{rem}}$, and the remnant's kick velocity $v_{\mathrm{kick}}$ in $\mathrm{km \ s}^{-1}$. Velocities are normalized to a binary system with total mass $M=10^6 \mathrm{M}_{\odot}$, and $\mathrm{M}_6 \equiv M/10^6 \ \mathrm{M}_{\odot}$.}\label{tab:initdata}
 \begin{tabular}{lccccccccc}
\hline\hline \\[-1.6ex]
  Run & $a_0 \ [M]$ & $p_{x}$ & $p_{y}$ &  $\hat{a}_{1}$ & $\hat{a}_{2}$ & $t_{\mathrm{Merger}} \ [M]$ & M$_{\mathrm{rem}}$ & $a_{z, \mathrm{rem}}$ & $v_{\mathrm{kick}} \ [\mathrm{km \ s}^{-1} \ M_6]$ \vspace{0.2pt}\\
\hline
  \texttt{UU} &   & 4.62e-4 & 8.19e-2 & (0.0, \ \ 0.0, \ \ 0.6) & (0.0, \ \ 0.0, \ \ 0.6) &  2529  &  0.951   &  0.858   & \textendash \\
\cmidrule(lr) {1-1}\cmidrule(lr) {3-10}
  \texttt{UD}  &   & 5.16e-4  &  8.43e-2 & (0.0, \ \ 0.0, \ \ -0.6) & (0.0, \ \ 0.0, \ \ 0.6) &  1989  &  0.951   &  0.622  &  261 \\
\cmidrule(lr) {1-1}\cmidrule(lr) {3-10}
  \texttt{DD}  & 12.162  & 6.24e-4 &  8.73e-2 & (0.0, \ \ 0.0, \ \ -0.6) & (0.0, \ \ 0.0, \ \ -0.6) &  1452  & 0.964  &  0.459 &  \textendash    \\
 \cmidrule(lr) {1-1}\cmidrule(lr) {3-10}
  \texttt{UUMIS}  &   & 4.6e-4 &  8.24e-2 & (-0.42, 0.0,  0.42) & (0.42, 0.0, 0.42) &  2401  &  0.935   &  0.811  & 1739 \\
\cmidrule(lr) {1-1}\cmidrule(lr) {3-10}
  \texttt{UDMIS}  &   & 5.16e-4 &  8.43e-2 & (-0.42, 0.0,  -0.42) & (0.42, 0.0, 0.42) &  2008  & 0.951  &  0.688  & 797 \\
\hline\hline
 \end{tabular}
\end{table*}
\vspace{.2cm}
In this Letter, we present the first GRMHD simulations of spinning binaries of black holes with spins either aligned, anti-aligned or misaligned with the orbital angular momentum, investigating how the spin inclination modifies the magnetohydrodynamical behavior of the plasma during the binary late-inspiral and merger.
As in \citeAliasTwo{Cattorini-2021}{Paper I}, the simulations presented here aim at extracting physically relevant information which can help improving our understanding of hot accretion flows onto merging MBHBs. Also, we explore the effects of individual spin orientation in a binary system, in order to determine how it affects the accretion flow and the properties of the emitted jet.
\section{Numerical Methods} \label{sec:num}
The details of the numerical setup employed for running the simulation presented here are thoroughly discussed in \citeAliasTwo{Cattorini-2021}{Paper I}. Below, we limit ourselves to a rapid overview of the mathematical and computational techniques adopted in the present work. Throughout the paper, we use geometrized units where $G=c=1$.
\subsection{Spacetime and Matter Fields Evolution}\label{sec:fieldsevo}
The simulations presented in this Letter are built upon the same configuration presented in \citeAliasTwo{Cattorini-2021}{Paper I}. They were run with the \texttt{Einstein Toolkit}\footnote{\url{http://einsteintoolkit.org}}\footnote{The Einstein Toolkit ``Turing'' release: \url{https://zenodo.org/record/3866075\#.YeGMu_so_QU}} framework \cite[][]{Loffler-2012}, using the ``moving puncture'' formalism \cite[][]{Campanelli-2006, vanMeter-2006}, and performing full GR evolution of both the spacetime metric and  the magnetohydrodynamic fields. The spacetime metric is evolved via the \texttt{McLachlan} \cite[][]{Husa-2006, Brown-2009} thorn using the BSSNOK formulation \cite[see, e.g., Refs.][]{Shibata-Nakamura-1995, Baumgarte-Shapiro-1998}. The MHD equations are solved in a flux-conservative formulation by the \texttt{IllinoisGRMHD} thorn \cite[][]{Noble-2006, Etienne-2015}. The divergence-free property of the magnetic field is guaranteed with the evolution of the vector and scalar potentials in the ``generalized'' Lorenz gauge, and the MHD equations are solved in the ideal MHD limit, i.e., we consider a perfectly conducting medium in which Maxwell's equations reduce to $\nabla_{\nu} F^{* \mu \nu}=0$.
All binary evolutions are carried out on adaptive-mesh refinement (AMR) grids provided by the \texttt{Carpet} driver \cite[][]{Schnetter-2004}.
We calculate the gas accretion rate using the thorn \texttt{Outflow} \cite[][]{Outflow-thorn}, which measures the rest-mass density flow across the BH apparent horizons.
The gravitational radiation generated during the late-inspiral, merger and ringdown is computed with the \texttt{WeylScal4} thorn via the Weyl curvature scalar $\Psi_4$, calculated given the fiducial tetrad of  \cite{Baker-2002}.
\begin{figure*}
\begin{center} 
\includegraphics[width=\textwidth]{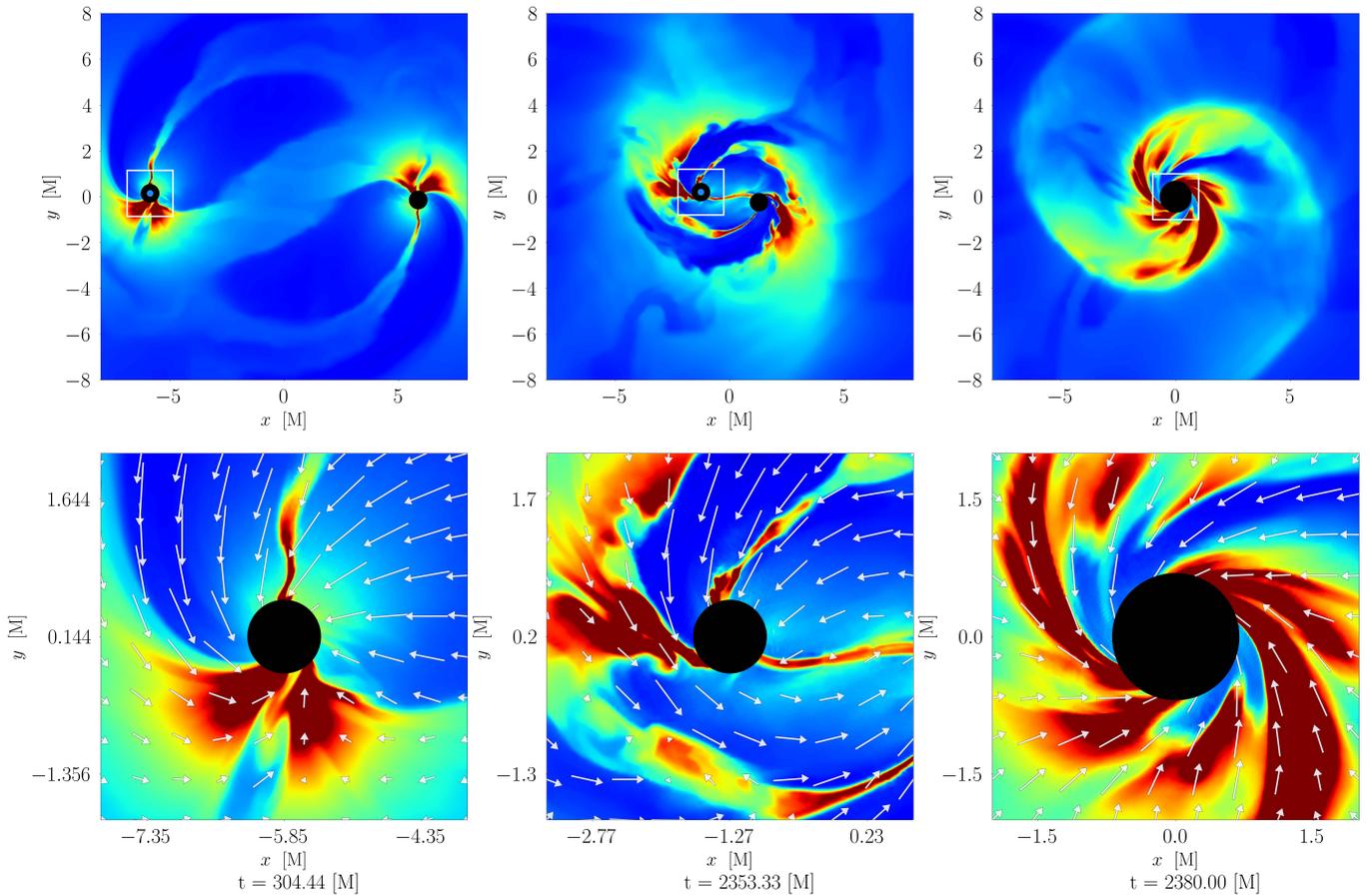}
\end{center}
\caption{(Top row): evolution of the rest-mass density $\rho$ (normalized to its initial value $\rho_0$) on the $xy$-plane for \texttt{UUMIS} configuration ($\hat{a}_1=(-0.42,0,+0.42)$, $\hat{a}_2=(+0.42,0,+0.42)$, $\beta_0^{-1}=0.31$); snapshots were taken, respectively, after $\sim$1 orbit (left), after $\sim$11 orbits (center), and at the time of merger (right).
(Bottom row): close-ups at the same epochs of the top panels of the region in the vicinity of the first black hole's apparent horizon (left and middle panels) and of the remnant black hole (right panel). Arrows denote velocity vectors. The black circles denote the interior of the apparent horizons. In the top panel a colored dot was added inside the BH with initial spin $\hat{a}_1$.}\label{fig:pr72rhoEvo}
\end{figure*}
\subsection{Initial Conditions}\label{subsec:num02}
The initial data of the simulations presented here were chosen consistently with \citeAliasTwo{Cattorini-2021}{Paper I}.
Since matter accretion tends to equalize the binary component masses, we have chosen to evolve equal-mass systems \cite[][]{Farris-2014a, Duffell-2020}.
We perform a set of five runs, (see Table \ref{tab:initdata}).  Each run evolves an equal-mass binary immersed in a uniform polytropic fluid ($p_0=\kappa\rho_0^{\Gamma}$, with $\rho_0=1$, $\kappa=0.2$, and $\Gamma=4/3$), which is threaded by an initially uniform magnetic field aligned to the orbital angular momentum. The gas is initially at rest with respect to the binary. The total mass of the system in code units is set to be $M=1$; the mass of each BH is $M/2$, and we assume the total mass of the gas to be negligible (i.e., we evolve Einstein's field equations in vacuum). The initial value of the magnetic field is chosen so that the initial magnetic-to-gas pressure is $\beta^{-1}_0 \equiv p_{\mathrm{mag,0}}/p_{\mathrm{gas,0}}=0.31$.

All simulations employ 11 refinement levels, with a resolution of 1/56 M on the finest grid, covering the apparent horizon radius of each BH with $\sim$ 20 grid points.
In accordance with \citeAliasTwo{Cattorini-2021}{Paper I}, the binaries are initialized on quasi-circular orbits at an initial separation $a_0 \approx 12 \ M$. 
The adimensional spins in each run have the same magnitude $a_1 = a_2 = 0.6$. In configuration \texttt{UU} (\texttt{DD)}, both spins are aligned (anti-aligned) with the orbital angular momentum $L_{\mathrm{orb}}$; in configuration \texttt{UD}, one spin is aligned, and the other is anti-aligned with $L_{\mathrm{orb}}$; in configurations \texttt{UUMIS} and \texttt{UDMIS}, the spins are misaligned with $L_{\mathrm{orb}}$.
In Table \ref{tab:initdata}, we display the initial puncture positions, momenta, and spins of our five runs, along with merger times $t_{\mathrm{merger}}$,  remnant spins $a_{\mathrm{rem}}$, and kick velocities $v_{\mathrm{kick}}$ (when present).
\section{Results} \label{sec:res}
To investigate the effects of spin orientation on the accretion flows, we study the evolution of the rest-mass density, velocity and magnetic-to-gas pressure fields, the mass accretion rate onto the BH horizons, and the emitted Poynting luminosity as diagnostics.
\begin{figure*}
\begin{center} 
\includegraphics[width=\textwidth]{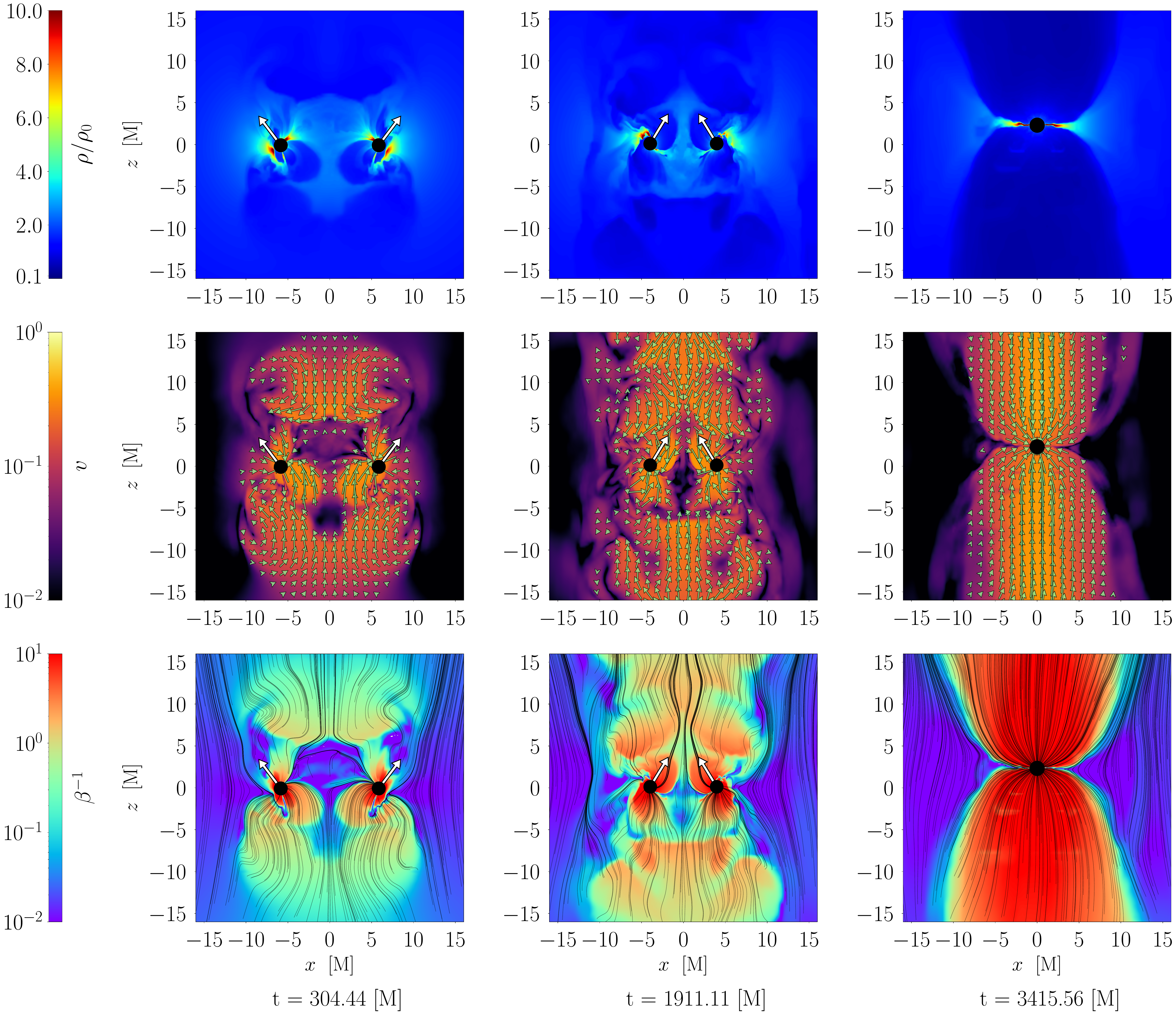}
\end{center}
\caption{(Top row): evolution of the rest-mass density $\rho$ (normalized to its initial value $\rho_0$) on the $xz$-plane.
(Middle row): evolution of the fluid velocity field $\vec{v}$ on the $xz$-plane. The colorbar refers to the magnitude $v=(v_x^2+v_z^2)^{1/2}$.  Arrows refer to the gas velocity field.
(Bottom row): evolution of the magnetic field lines on the $xz$-plane. The colorbar refers to the magnetic-to-gas pressure ratio $\beta^{-1}$. The regions inside the BH horizons have been masked out. 
All snapshots refer to our \texttt{UUMIS} configuration ($\hat{a}_1=(-0.42,0,+0.42)$, $\hat{a}_2=(+0.42,0,+0.42)$, $\beta_0^{-1}=0.31$); snapshots were taken, respectively, after $\sim$1 orbit (left column), after $\sim$8 orbits (middle column), and at a time equal to $\sim$1000 $M$ after merger (right column). The white arrows over BHs in the left and middle column denote spin vectors.}\label{fig:pr72Evo}
\end{figure*}
\subsection{MHD fields evolution} \label{subsec:rho}
In the top row of Fig. \ref{fig:pr72rhoEvo}, we display the evolution of the rest-mass density $\rho$ (normalized to its initial value $\rho_0$) on the $xy$-plane for \texttt{UUMIS} configuration ($\hat{a}_1=(-0.42,0,+0.42)$, $\hat{a}_2=(+0.42,0,+0.42)$). We choose \texttt{UUMIS} configuration as our representative model, for it is a clear demonstration of the effects of spin on the accretion rate and Poynting luminosity (see Sec. \ref{subsec:mod}). A colored dot was added inside the BH with initial spin $\hat{a}_1$.
On the bottom row, we show close-ups of the region in the vicinity of the first black hole's apparent horizon (left and middle panels), and of the remnant black hole (right panel). White arrows denote the velocity vectors of the fluid.

The behavior of the plasma in the equatorial plane resembles the evolution of the magnetized simulations of \cite{Giacomazzo-2012}, \cite{Kelly-2017}, and \citeAliasTwo{Cattorini-2021}{Paper I}. 
The main differences with respect to those configurations appear in the fields evolution in the polar plane.
In Fig. \ref{fig:pr72Evo}, we display 2D slices in the $xz$-plane representing the evolution of the rest-mass density field $\rho/\rho_0$ (top row), velocity field (middle row), and magnetic-to-gas pressure field $\beta^{-1}$ (bottom row) for our \texttt{UUMIS} run. 
\begin{figure*}
\begin{center} 
\includegraphics[width=.9\textwidth]{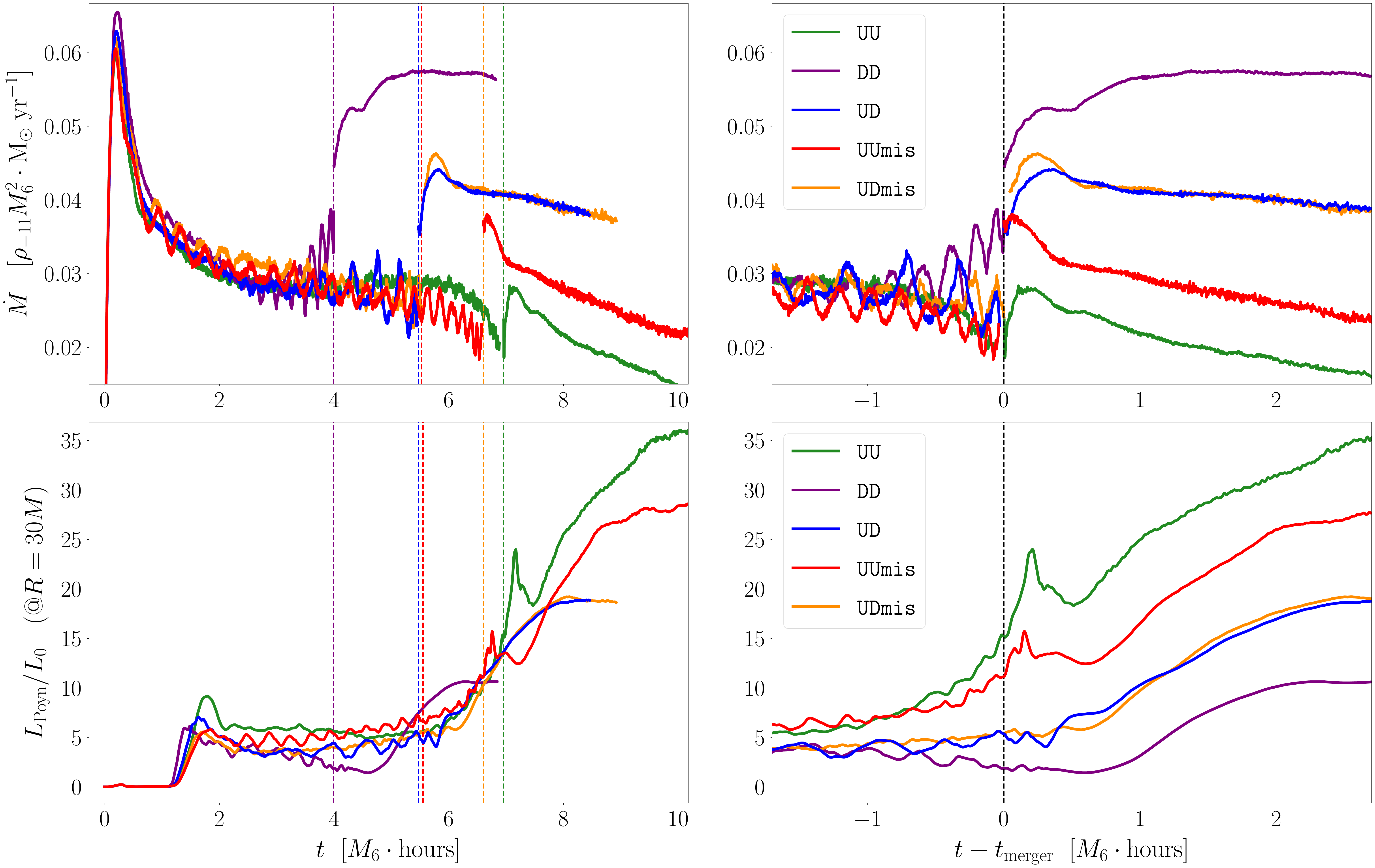}
\end{center}
\caption{(Top row, left): time-dependent accretion rates $\dot{M}$ in units of solar masses per year for the five configurations. The magnitudes of $\dot{M}$ are scaled to a binary of total mass $10^6 \ \mathrm{M}_{\odot}$ and a gas with initial uniform rest-mass density $\rho=10^{-11} \ \mathrm{g \ cm}^{-3}$. The pre-merger accretion rate is calculated onto both BH horizons; the post-merger accretion rate is calculated onto the remnant Kerr BH horizon. The vertical dotted lines mark the merger times, and colors highlight the different configurations. (Top row, right): comparison of $\dot{M}$ in the late-inspiral, merger and post-merger stages; time is rescaled with the merger time $t_{\mathrm {merger}}$. 
(Bottom row, left): time evolution of the Poynting luminosity $L_{\mathrm{Poynt}}$ for the five configurations. The luminosity is extracted on a sphere of radius $R_{\mathrm{ext}} = 30 \ M$ centered in the origin of the system. The values of $L_{\textrm{Poynt}}$ are in units of $L_0 \equiv 2.347\times 10^{43} \rho_{-11}M_6^2$ erg s$^{-1}$, with $\mathrm{M}_6 \equiv M/10^6 \ \mathrm{M}_{\odot}$, and $\rho_{-11} \equiv \rho/10^{11} \ \mathrm{g \ cm}^{-3}$. (Bottom row, right): comparison of $L_{\mathrm{Poynt}}$ in the late-inspiral, merger and post-merger stages.}\label{fig:mdotLP}
\end{figure*}
The simulation begins $\sim$2400 $M$ ($\sim$11 orbits) before merger, with an initially uniform gas and a uniform magnetic field directed along the orbital axis. After a time as short as one orbit, matter starts to concentrate around each BH, forming two overdensities distributed in planar ``disklike'' structures (see also \cite{Giacomazzo-2012, Kelly-2017} and Fig.1 in \citeAliasTwo{Cattorini-2021}{Paper I}). However, unlike previous results of non-spinning and aligned-spin BHBs, the flow structure around the BHs is approximately orthogonal to the spin axes (Fig. \ref{fig:pr72Evo} first row, left and central panels). These structures persist until merger, when matter settles into a disklike overdensity around the Kerr BH remnant which weakly oscillates around the orbital plane (Fig. \ref{fig:pr72Evo} first row, right panel), and is dragged by the BH remnant, which is recoiling with a velocity $v_{\mathrm{kick}} \approx 1700$ $M_6$ km s$^{-1}$ in the direction of the $z$-axis.
 
The inclination of the disklike overdensities can be understood in terms of the magnetic field behavior in the proximity of the horizons: as the BHs orbit around each other, the magnetic field lines accumulate near the horizons, and are oriented towards the spin axis of each BH (Fig. \ref{fig:pr72Evo} third row, left and central panels) leading to the formation of magnetically dominated funnels, which we call ``protojets''\footnote{We define protojets as magnetically dominated regions with a strong, localized Poynting flux, in which the net fluid flow is directed inward.} following \cite{Kelly-2021}. We find that the smaller protojets emerging from the individual Kerr BHs are always oriented toward the BHs spin direction at distances  $\lesssim 5\,M$, whereas, at larger radii, the protojets are aligned to the orbital ($z$-)axis. This effect is in accordance with the results of simulations of individual post-merger Kerr BHs by \cite{Kelly-2021}.
These regions eventually merge and form a protojet collimated in the polar direction (Fig. \ref{fig:pr72Evo} third row, right panel). 

\subsection{Mass Accretion Rate}\label{subsec:mdot}
The rest-mass flux across the horizons of each BH is computed by the \texttt{Outflow} thorn \cite[][]{Outflow-thorn}.
In the top panels of Fig. \ref{fig:mdotLP}, we display the time evolution of the mass accretion rate $\dot{M}$ for our five configurations. The values of $\dot{M}$ are in units of solar masses per year, normalized for a binary system of total mass $10^6 \ \mathrm{M}_{\odot}$ immersed in a plasma with initial density $10^{-11} \ \mathrm{g \ cm}^{-3}$. On the left panel, the dashed colored lines mark the time of merger for each configuration. On the right panel, we compare the accretion rates  of the five runs in the late-inspiral, merger and post-merger stages; the time axis is aligned by the merger time of the binary $t_{\mathrm{merger}}$.
\begin{figure*}
\begin{center} 
\includegraphics[width=\textwidth]{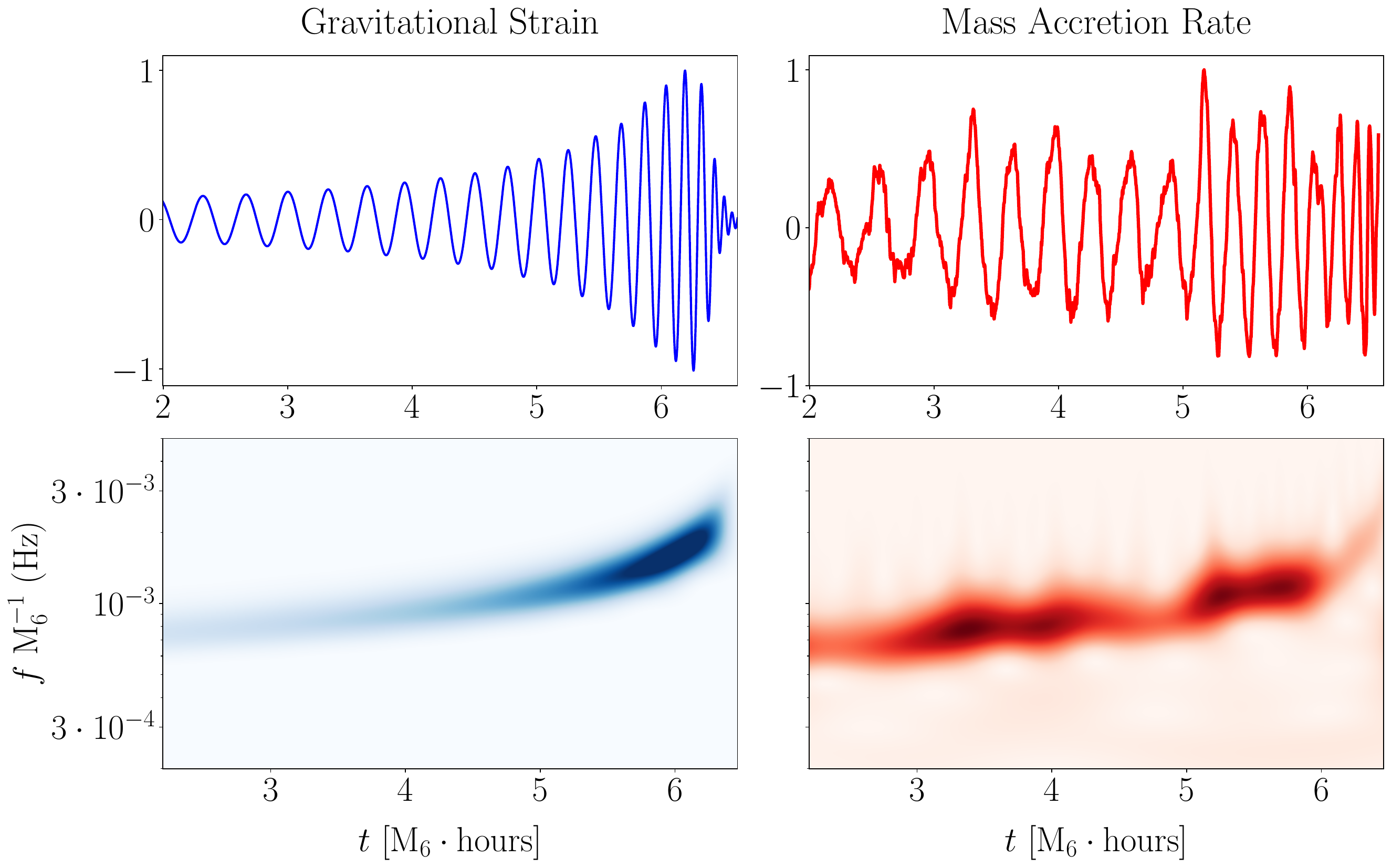}
\end{center}
\caption{(Top row, left): GW strain of \texttt{UUMIS} run extracted via the Weyl scalar $\Psi_4$ with the \texttt{WeylScal4} thorn and normalized to its maximum value at merger. (Top row, right): time-dependent pre-merger mass accretion rate $\dot{M}^* = \dot{M}-p_6$ over both BH horizons for \texttt{UUMIS} run, also normalized to its maximum value.
(Bottom row, left): time-frequency representation of the GW strain via wavelet PSD, showing the frequency increase of the signal over time. (Bottom row, right): time-frequency representation of $\dot{M}^*$ via wavelet PSD, showing similarity with the GW strain in the frequency increase over time. Time and frequency units are normalized to a binary system with total mass $M=10^6 \mathrm{M}_{\odot}$, and $\mathrm{M}_6 \equiv M/10^6 \ \mathrm{M}_{\odot}$.} \label{fig:wavelets}
\end{figure*}

In \citeAliasTwo{Cattorini-2021}{Paper I}, we have shown that higher values of the spin parameter result in a suppressing effect on $\dot{M}$. Analogous behavior is observed in our current set of simulations: run \texttt{UU} (which results in a Kerr BH remnant with spin parameter $a_{\mathrm{rem}} \sim 0.86$) has the lowest post-merger accretion rate ($\dot{M}_{\texttt{UU}} \sim 0.15 \ \rho_{-11}M^2_6 \ \mathrm{M}_{\odot} \mathrm{yr}^{-1}$), whereas run \texttt{DD} ($a_{\mathrm{rem}} \sim 0.46$) reaches the highest post-merger accretion rate ($\dot{M}_{\texttt{DD}} \sim 0.6 \ \rho_{-11}M^2_6 \ \mathrm{M}_{\odot} \mathrm{yr}^{-1}$). Other configurations exhibit in-between values of $\dot{M}$. In general, a higher post-merger spin results in a lower accretion rate.

\subsection{Poynting Luminosity}\label{subsec:lpoyn}
In \citeAliasTwo{Cattorini-2021}{Paper I}, we investigated how the features of the EM Poynting luminosity $L_{\mathrm{Poynt}}$ relate to the initial magnetic-to-gas pressure $\beta^{-1}_0$ and black holes' spins. In agreement with the results of \cite{Kelly-2017}, we observed that configurations with the same spin reach approximately the same value of $L_{\mathrm{Poynt}}$, regardless of $\beta^{-1}_0$. However, we found that the peak Poynting luminosity, which is reached shortly after merger, is particularly dependent on the magnitude of the remnant's spin parameter. We show in the bottom panels of Fig. \ref{fig:mdotLP} the time evolution of $L_{\mathrm{Poynt}}$ for our five runs. The values of $L_{\textrm{Poynt}}$ are expressed in units of $L_0 \equiv 2.347\times 10^{43} \rho_{-11}M_6^2$ erg s$^{-1}$ (see \citeAliasTwo{Cattorini-2021}{Paper I}).

The main features of the Poynting luminosity ``light curves'' for \texttt{UU} and \texttt{UUMIS} runs (see green and red curves in Fig. \ref{fig:mdotLP}, bottom panels) are similar to those displayed in \citeAliasTwo{Cattorini-2021}{Paper I}: an initial steep rise (i), followed by a slow growth stage (ii) across the binary inspiral; a peak (iii) corresponding to merger, and a rapid climb (iv) toward a steady value (v).
Configurations \texttt{UD} and \texttt{UDMIS} (blue and yellow curves, respectively) exhibit similar trends, except for the absence of evident peaks corresponding to binary mergers. Finally, configuration \texttt{DD} (purple curve) shows moderate decrease across the inspiral, and rapidly climbs over the merger and ringdown stages. As \texttt{UD} and \texttt{UDMIS} runs, also configuration \texttt{DD} does not feature a peak during merger.

We verified that the magnitude of the post-merger steady values of $L_{\mathrm{Poynt}}$ of our five configurations approximately scales with the spin parameter squared, $a^2$, in agreement with the Blandford-Znajek formula \cite[][]{BZ-1977}. This scaling will be subject of a future investigation involving a broader  family of spinning configurations.

\subsection{Modulations}\label{subsec:mod}
Over the last decade, a number of explorative works have reported that quasiperiodic features in the light curve of a MBHB system may arise thanks to the fueling-rate variability in mini-disks around each BH due to the periodic interaction of the BHs with the inner edge of the CBD \cite[][]{Noble-2012, Farris-2014a, Farris-2015a, Tang-2018}, or because of relativistic Doppler modulation \cite[][]{Haiman-2017}. 
In our work, quasiperiodic variability is found in the time-dependent accretion rates for the five configurations presented (see Fig. \ref{fig:mdotLP}, top-left panel); similar \textendash  though weaker\textendash \  oscillations are present also in the evolution of the Poynting luminosity (Fig. \ref{fig:mdotLP}, bottom-left panel), but only when it is extracted on spheres of radii 10 or 30 $M$; when $L_{\mathrm{Poynt}}$ is extracted at higher radii (in our cases, at 50, 80, and 100 $M$), its oscillatory behavior disappears and the light curves across the inspiral are smooth. This is in agreement with simulations of non-spinning BHs carried out by \cite{Kelly-2017} (see Fig. 21 therein).
In what follows, we focus on the time variability of $\dot{M}(t)$, since it provides a more immediate correlate to detectable EM emission via accretion luminosity.

We observe that, over the last $\sim$10 orbits of \texttt{UUMIS} configuration, the accretion rate displays a clear modulation, with oscillations occurring with amplitude $\sim$10\% times the average rate.
The amplitude of such oscillations is lowest for \texttt{UU} run ($\sim$1\%), and largest for \texttt{UD} and \texttt{DD} runs ($\sim$20\%).

In order to examine the harmonic connection of the pre-merger accretion rate with the chirping GW signal, we perform wavelet power spectral density (WPSD) analysis \cite[][]{Chatterji-2004}. We fit the pre-merger accretion rate with a 6th-order polynomial $p_6$, and subtract it to $\dot{M}$ in order to remove the initial transient and the late decrease prior to merger. In Fig. \ref{fig:wavelets} (top row), we plot the gravitational strain computed via the Weyl scalar $\Psi_4$ (left), and the quantity $\dot{M}^* = \dot{M}-p_6$ (right), representing the pre-merger accretion rate for \texttt{UUMIS} run, both normalized to their maximum value. To investigate the time-frequency behavior of $\dot{M}^*$, we compute its WPSD and compare it to the WPSD of the GW strain (Fig. \ref{fig:wavelets}, bottom row). Our main result is that the $\dot{M}^*$ time series (right panel) oscillates with a frequency increment that closely resembles the increase of characteristic GW chirp frequency with time (left panel).

The occurrence of modulations of the accretion rate on such short space- and time-scales is remarkable, for it can be possibly translated into quasiperiodic oscillations in the EM light curve, allowing for the identification of an EM counterpart to the GW event. In general, the quasiperiodic modulations appear to be independent of the larger-scale structure of the binary gaseous environment, and depend only upon the magnetohydrodynamic features of the accreting fluid near the horizons.

\vspace{.2cm}
The mechanism behind these fluctuations is not completely clear and will be subject of future investigation. Still, we verified that no modulation in $\dot{M}$ is present whenever the fluid is not initially threaded by a magnetic field, hence suggesting that such modulations arise as a result of the interplay of magnetic fields and strong, dynamical gravitational fields.

\section{Conclusions} \label{sec:con}

We have presented the first GRMHD simulations of equal-mass spinning black hole binary mergers with spins misaligned with respect to the orbital angular momentum.
We performed a suite of five simulations of BHBs initially immersed in a uniform gas cloud with uniform magnetic field aligned with the binary orbital angular momentum $L_{\mathrm{orb}}$.
Each configuration evolves BHs with spins of the same magnitude, but differing in the orientation relative to $L_{\mathrm{orb}}$.
In agreement with previous results from \citeAliasTwo{Cattorini-2021}{Paper I}, we found that a higher post-merger spin of the remnant BH corresponds to lower mass accretion rates onto the BH horizon, and that the post-merger value of the Poynting luminosity is proportional to the square of the spin parameter of the newly formed BH. We discovered the occurrence of quasiperiodic oscillations in the mass accretion rate - and, possibly, in the emitted EM radiation -  during the inspiral phase, and found that the oscillation frequency of the accretion rate increases with time to merger,  mimicking the gravitational chirp. This similarity seemingly arises as a consequence of the interplay of the magnetic field in the vicinity of a merging binary and the spins of the individual black holes. This finding suggests that quasiperiodicities in the pre-merger accretion rate are not exclusive of environments in which a BHB is embedded in a circumbinary accretion disk. Such oscillations may give rise to quasiperiodic EM emission in the X-ray band, which could potentially be detected by the future Athena mission \cite[][]{AL-synergies}, thus providing a useful signature of the EM signal concurrent to the gravitational emission of merging massive black hole binaries.

\begin{acknowledgements}
We thank Bernard Kelly for useful discussions. 
All simulations were performed on MARCONIA3 cluster at CINECA (Bologna,  Italy). The numerical calculations have been made  possible through a CINECA-INFN agreement, providing access to resources on MARCONIA3 (allocation INF21\_teongrav). M.C. and F.H. acknowledge funding from MIUR under the grant PRIN 2017-MB8AEZ.
\end{acknowledgements}
\bibliography{bibliography}{}
\bibliographystyle{aasjournal}

\end{document}